\pdfoutput=1

\documentclass{vgtc}                          

\usepackage{mathptmx}
\usepackage{graphicx}
\usepackage{times}

\usepackage[utf8]{inputenc}
\usepackage[T1]{fontenc}

\vgtccategory{Research}

\vgtcinsertpkg

\title{Perceiving Mass in Mixed Reality through\\ Pseudo-Haptic Rendering of Newton's Third Law}

\author{Paul Issartel\thanks{e-mail: \{paul.issartel,mehdi.ammi\}@limsi.fr}\\ %
        \parbox{1.4in}{\scriptsize \centering LIMSI-CNRS \\ Univ. Paris-Sud} %
\and Florimond Guéniat\thanks{e-mail: contact@gueniat.fr}\\ %
     \parbox{1.4in}{\scriptsize \centering Department of Mathematics \\ Florida State University} %
\and Sabine Coquillart\thanks{e-mail: sabine.coquillart@inria.fr}\\ %
     \parbox{1.4in}{\scriptsize \centering INRIA \\ Univ. Grenoble Alpes, LIG \\ CNRS, LIG} %
\and Mehdi Ammi\footnotemark[1]\\ %
     \parbox{1.4in}{\scriptsize \centering LIMSI-CNRS \\ Univ. Paris-Sud}}

\teaser{
  \begin{minipage}{\textwidth}
    \vspace*{-3mm}
    \centering
    \includegraphics[height=2.48cm]{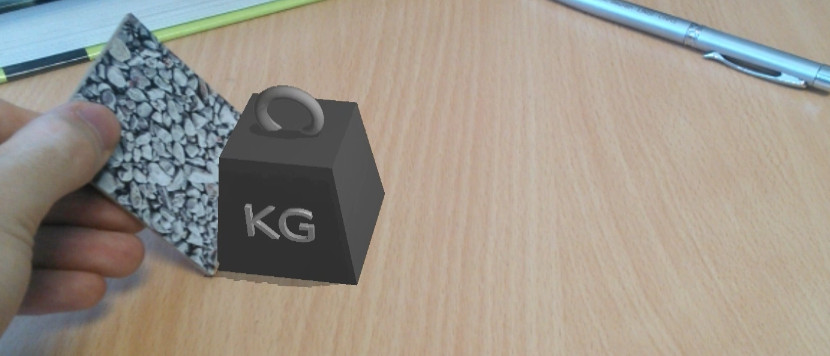}\hspace{.25mm}%
    \includegraphics[height=2.48cm]{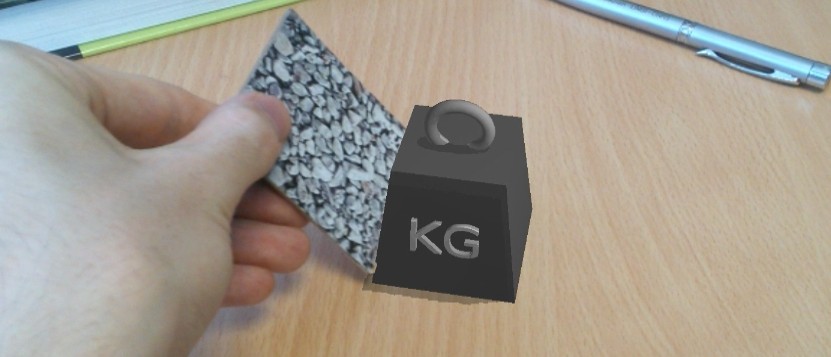}\hspace{.25mm}%
    \includegraphics[height=2.48cm]{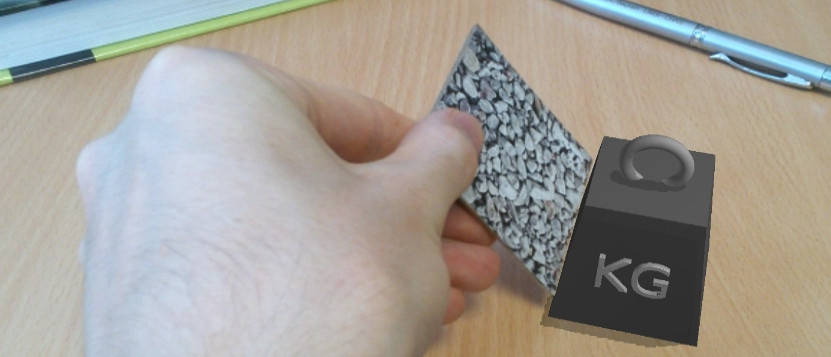}
    \\[.8mm]%
    \includegraphics[height=2.48cm]{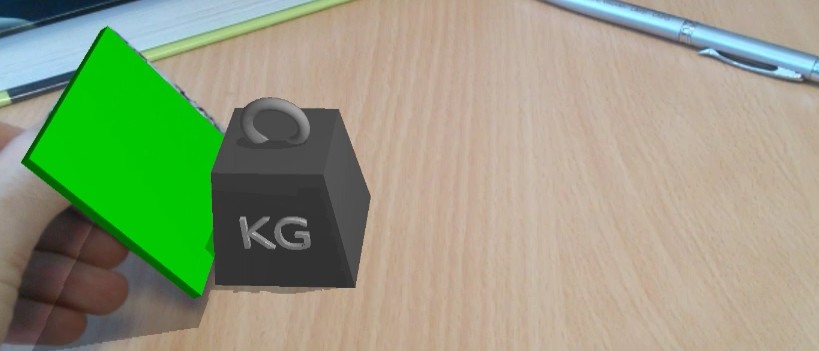}\hspace{.25mm}%
    \includegraphics[height=2.48cm]{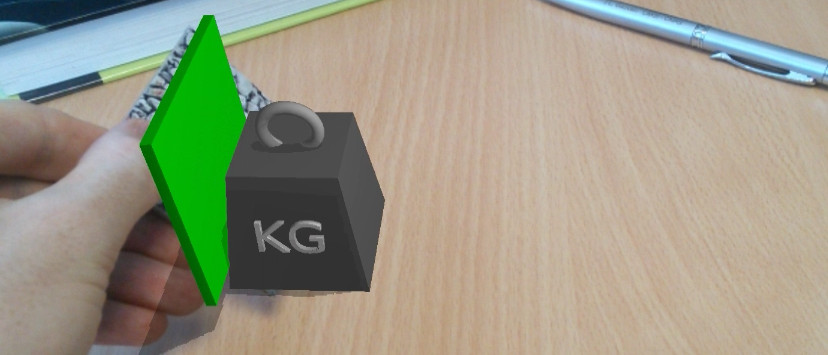}\hspace{.25mm}%
    \includegraphics[height=2.48cm]{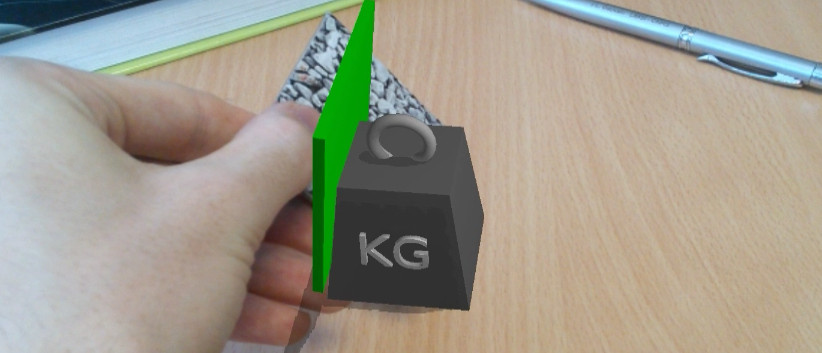}
  \end{minipage}%
  \caption{%
    In the top row, a real-world effector object manipulated by the user does not encounter any reaction force when pushing a virtual object, which appears to be massless. In the bottom row, the effector is visually affected by a reaction force, and the virtual mass becomes perceivable.}
  \label{fig:teaser}
  \vspace{-4.6mm}
}

\abstract{%
  In mixed reality, real objects can be used to interact with virtual objects. However, unlike in the real world, real objects do not encounter any opposite reaction force when pushing against virtual objects. The lack of reaction force during manipulation prevents users from perceiving the mass of virtual objects. Although this could be addressed by equipping real objects with force-feedback devices, such a solution remains complex and impractical.

  In this work, we present a technique to produce an illusion of mass without any active force-feedback mechanism. This is achieved by simulating the effects of this reaction force in a purely visual way. A first study demonstrates that our technique indeed allows users to differentiate light virtual objects from heavy virtual objects. In addition, it shows that the illusion is immediately effective, with no prior training. In a second study, we measure the lowest mass difference (JND) that can be perceived with this technique. The effectiveness and ease of implementation of our solution provides an opportunity to enhance mixed reality interaction at no \mbox{additional cost}.
} 


\CCScatlist{
  \CCScat{H.5.1}%
  {Information Interfaces And Presentation}%
  {Multimedia Information Systems}%
  {Artificial, augmented, and virtual realities}
}

\usepackage[protrusion,expansion,tracking,kerning,spacing]{microtype}
\microtypecontext{spacing=nonfrench}
\SetTracking{encoding={*}, shape=sc}{0}
\usepackage{amsmath}
\usepackage{color}
\usepackage{enumitem}
\setitemize{noitemsep,topsep=1mm}
\setenumerate{noitemsep,topsep=1mm}
\setdescription{noitemsep,topsep=1mm}
\usepackage{textcomp}
\usepackage{siunitx}
\usepackage{array}
\usepackage{booktabs}
\usepackage{url}

\begin{document}


\firstsection{Introduction}

\maketitle

Mixed reality involves the merging of real and virtual worlds~\cite{milgram94} so that real objects and virtual objects appear to co-exist and interact with each other. This interaction can be visual, such as occlusion and~shadowing between objects, which enhances the illusion that virtual objects are present alongside real objects. Another way to increase this illusion is through realistic physical interaction~\mbox{\cite{buchanan08,piumsomboon11}}. Virtual objects become affected by physical laws, and react to collisions with other objects. In particular, they can be pushed by real objects manipulated by the user.

However, this interaction remains one-way. Even though virtual objects react to collisions with real objects, the reverse is not true. Virtual objects do not physically exist, so they cannot prevent a real object from passing through them. Therefore, the user does not feel any reaction force when pushing a real object against a virtual object.

This lack of reaction presents an issue for mass perception. As per Newton's third law~\cite{newton87}, an object pushing against another should experience an equal reaction force in the opposite direction. Heavier objects have more momentum than light objects, so they require a greater force to be moved. A strong reaction force would thus normally indicate that the object being pushed is heavy. If the user does not feel any forces from virtual objects, then the objects' masses become \mbox{unperceivable}~(Figure~\ref{fig:teaser}).

One way to address this issue is to equip manipulable objects with force-feedback devices~\cite{jeon11}. But it requires complex devices, such as robotic haptic arms, in order to produce a reaction force. It also does not scale well to the large number of potentially manipulable objects in the real world. Thus, using active force-feedback devices remains difficult and impractical.

In this paper, we propose a new solution to recreate the missing reaction force predicted by Newton's third law. Our technique exploits the fact that the visual modality alone is powerful enough to induce illusory haptic sensations. This is known as ``pseudo-haptic feedback''~\cite{lecuyer00}. As explained above, it is difficult to apply a true reaction force to a real object when it hits a virtual object. However, pseudo-haptics can be used to \emph{visually simulate} the effects this force would have on the real object. Our technique consists in displaying a clone of the object that \emph{appears} to be pushed back on contact, even though the real object itself is not actually affected. By visually reproducing the effects of the missing reaction force, our proposed technique thus allows users to perceive the mass of virtual objects \mbox{without any active force-feedback.}

\section{Related work}

\subsection{Physics-based interaction}

An increasing number of mixed reality applications are using \mbox{realistic} physics, to enhance the user experience and support natural interaction with virtual content. Buchanan~et~al.~\cite{buchanan08} presented an educational augmented reality game, where virtual objects could be manipulated through markers and interact with each other in a physically correct way. There was no physical interaction between real and virtual objects though. Wilson~\cite{wilson07} and Piumsomboon~et~al.~\cite{piumsomboon11} used depth sensors to scan the environment in real-time, allowing virtual objects to react physically to real objects. However, real objects were still not affected by virtual objects. From the physical simulation point of view, they were considered as immovable. Kim~et~al.~\cite{kim11} discussed the issues caused by immovable real-world objects. They demonstrate that it leads to unrealistic physical behavior, which they call ``physical artifacts''. In particular, one of these artifacts is that real objects do not experience an opposite reaction force when they collide with virtual objects, and thus are not pushed back. As we explained above, this makes light and heavy virtual objects indistinguishable.

\subsection{Active force feedback}

One way to allow real objects to be affected by virtual forces is to equip them with actuators. For instance, Kang~and~Woo~\cite{kang11} proposed \emph{ARMate}, an application where a real-world toy cart can be pushed by a virtual character. To achieve this, the cart is equipped with motors and a wireless controller. Although technically possible, equipping all potentially movable real objects with actuators would not really be practical. Another way to apply forces to real objects is to attach them to external force-feedback devices. Jeon~and~al.~\cite{jeon11} showed how a robotic haptic arm can be used to modulate forces experienced by an effector object. They were able to alter the subjects' perception of weight when lifting the object. The SPIDAR~\cite{kim00} is another device that can be used for this purpose. With this device, subjects were able to differentiate the weight of 3~different virtual objects. A~major drawback of this solution is that force-feedback devices are expensive, bulky and hardly transparent to the user. Furthermore, this solution can only be applied to a single object at the same time. Adding force-feedback to multiple real-world objects would require multiple devices. A different approach was proposed by Minamizawa~et~al.~\cite{minamizawa07}. Their haptic glove can generate a sensation of force by applying pressure and shearing to the skin. This would allow the user to perceive, at least partially, a reaction force when using any real object to push against a virtual object. However, this solution requires a special glove which has to be worn by the user.

\subsection{Pseudo haptic feedback}

Since adding active force feedback to real objects is technically difficult, it would be more convenient to simulate it by other means. A typical mixed reality system is capable of altering the visual representation of the real world in order to display virtual objects, and the visual modality is known to dominate the haptic modality for spatial interaction tasks~\cite{hatwell03}. Thus, in the absence of true haptic forces, visual dominance may be used to simulate force feedback.

An example that illustrates the influence of vision is the size-weight illusion~\cite{murray99}. When subjects are presented with two objects of equal mass but different sizes, they generally perceive the smaller object as heavier. Another example is the material-weight illusion: Buckingham~et~al.~\cite{buckingham09} showed that visual cues about \mbox{material} (polystyrene, wood, aluminum...) can influence the perceived weight of an object. These illusions suggest that the visual modality can indeed be used to simulate mass differences between virtual objects. However, altering the size or texture of objects is not a viable solution in mixed reality since virtual objects are primarily defined by their visual appearance.

An alternative is to visually change the physical behavior of manipulated objects. It has been shown that differences between the real and apparent motion of a manipulated object can produce illusory forces~\cite{lecuyer09}. This illusion is called \emph{pseudo-haptic feedback}. In an experiment called the ``swamp'', Lécuyer~et~al.~\cite{lecuyer00} showed that reducing the control-display ratio of a passive input device produces an illusion of viscosity on the controlled object. In a second experiment, they showed that the same approach can simulate stiffness. Crison~et~al.~\cite{crison04} were able to simulate material resistance in a virtual milling application by visually changing the apparent speed of the tool. Dominjon~et~al.~\cite{dominjon05} showed that control-display ratio can be used to influence the perceived mass (through weight or inertia) of otherwise identical \mbox{objects}.

The previous examples were not set in mixed reality. Altering the control-display ratio of a real object is much more challenging in mixed reality, since the user can see movements of the real object. Pusch~et~al.~\cite{pusch08} proposed a technique to generate pseudo-haptic feedback in augmented reality. They achieve this by displacing the entire image of the user's hand, in order to recreate the sensation of pressure exerted by a virtual flow. Taima~et~al.~\cite{taima14} presented a similar technique that visually amplify the movement of the user's hand during manipulation of a real object. They were able to influence the perceived weight of the object, and to influence user fatigue. These techniques are very powerful since they give full control on the perceived movement of any real-world object. However, they are also technically difficult to implement. Both projects cited above had to rely on a blue background to extract images of the hand and the manipulated objects. They are also restricted to immersive video see-through HMDs. Optical see-through and non-immersive devices would not work, as users would still be able to see their real hands.

In this work, we present a technique to reproduce the reaction force that should occur when a real object encounters a virtual object. This reaction force will then allow users to perceive the mass of virtual objects. Our technique avoids the constraints of active force-feedback devices by using pseudo-haptics. But rather than displacing entire parts of the visual environment, as it has been done before, we take a much simpler approach.

\section{Visually decoupling real objects from their\\ virtual representation}
\label{sec:decoupling}

\begin{figure}[h]
  \centering
  \vspace{-1.5mm}
  \includegraphics[height=2.75cm]{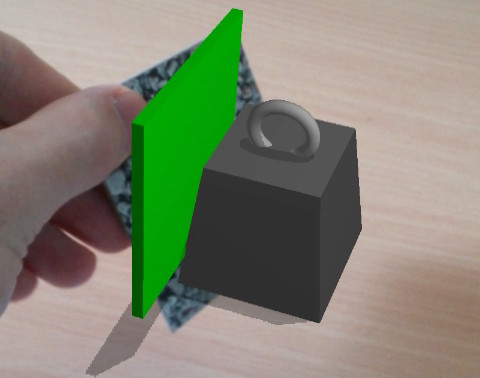}
  \hspace{1mm}
  \includegraphics[height=2.75cm]{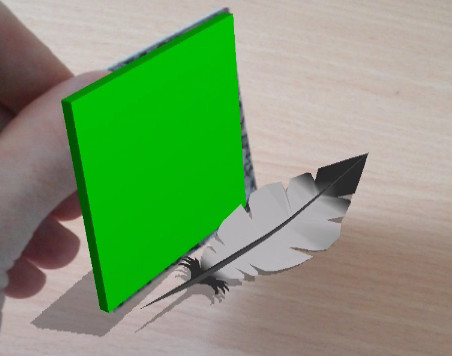}
  \vspace{-1.25mm}
  \caption{The virtual clone is visually decoupled from the position and orientation of the real object. It has a contrasting color to increase the visibility of decoupling. The strength of this decoupling effect depends on the mass of the virtual object being pushed. }
  \label{fig:decoupling}
  \vspace{-1.75mm}
\end{figure}

Our technique consists in creating a virtual clone of a real object, with the same size and shape, and attaching it to the real object with a physical constraint. This clone will act as a physical proxy for the real object. Since the clone is virtual, we are now able to control it entirely. It can be made an integral part of the physical simulation and interact properly with other virtual objects. Specifically, it can be \emph{pushed back} when encountering another virtual object, as predicted by Newton's third law. The strength of this effect will depend on the mass of the encountered virtual object. Light objects will barely affect the clone and will be strongly pushed back, whereas heavy objects will strongly affect the clone and barely move themselves~(Figure~\ref{fig:decoupling}). Thus, the displacement of the clone makes it possible to perceive the masses of virtual objects.

However, the virtual clone should not be pushed too far from its real counterpart. Otherwise, the two objects would appear as separate. It should stay as close as possible to the real object, but still be affected by virtual forces. This is accomplished by linking them with a 6-DOF spring constraint. Under such a constraint, motion along each degree of freedom is constrained by a spring (Figure~\ref{fig:spring-constraint}). The spring constraint tends to bring back the virtual clone to the position and orientation of the real object, with a force that increases with the distance. Therefore, the virtual clone stays attached to the real object, but is also able to deviate from the real position and orientation when pushed by virtual objects.

By creating a virtual clone attached to a real object, and making it affected by the reaction force during contact with virtual objects, we produce a pseudo-haptic illusion of the reaction force without any active force-feedback. Our technique does not require to visually displace the real object, making it much easier to implement than previous immersive approaches. Since the real view is not altered, two versions of the same object are visible (the real one and the clone). The spring constraint ensures they remain close enough to be considered as a single entity. Our technique loosens the coupling between the physical representation of the real object, materialized by its virtual clone, and the real object itself. They are linked, but not strictly coupled. In the rest of this paper, we will thus call this technique the \emph{decoupling technique}.

\section{Implementation}
\label{sec:implementation}

One of the main advantages of our proposed decoupling technique is that it is only based on visual rendering and does not require additional hardware. The software implementation is also straightforward, as it simply consists in simulating a 6-DOF spring constraint --~\mbox{assuming} that a physics engine is available to compute the remaining collision detection and rigid body dynamics. In our implementation, we used the Bullet\footnote{\url{http://www.bulletphysics.org/}} physics engine for this purpose.

\subsection{Spring constraint model}

\begin{figure}[h]
  \centering
  \includegraphics[height=3cm]{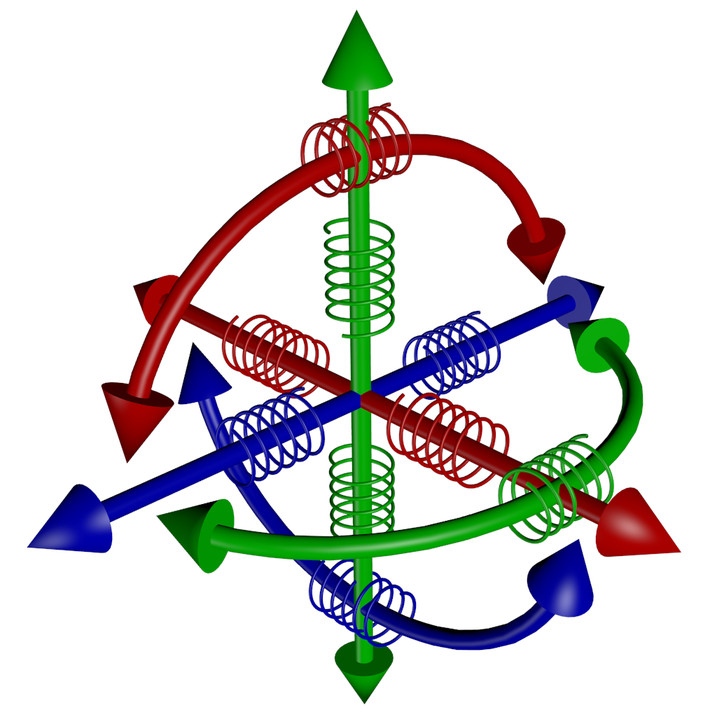}
  \caption{Schematic representation of a 6-DOF spring constraint.}
  \label{fig:spring-constraint}
\end{figure}

A 6-DOF spring constraint can be modelled as a spring-mass system for each 6~degrees of freedom (Figure~\ref{fig:spring-constraint}), where the virtual clone (``mass'') is attached to the real object with a spring. These springs are either extension springs, on translation axes, or torsion springs, on rotation axes. Each spring has a \emph{stiffness} parameter. As will be seen later, stiffness is of crucial importance in the effectiveness of this technique for mass perception.

However, plain springs would lead to instabilities when virtual forces stop being applied to the virtual clone. The clone would oscillate around the position and orientation of the real object, which would be puzzling to the user. To solve this issue, we use a \mbox{viscoelastic} constraint in which a damping component slows down the virtual clone when returning to its base configuration. The damping coefficient should be small enough to let the clone return as quickly as possible to the real object, but it should also be large enough to prevent oscillations. The exact value that satisfies both constraints is called the \emph{critical} damping coefficient.

The spring constraint is simulated by continuously applying forces and torques to the virtual clone, that attempt to bring it back to the real position and orientation. Per Hooke's law, the force exerted by a spring is equal to $-k\vec{d}$, where~$k$ is the spring stiffness and~$\vec{d}$ is the displacement. The damping force is equal to $-c\vec{v}$, where~$c$ is the damping coefficient and~$\vec{v}$ is the velocity of the clone relative to the real object. The force~$\vec{F}$, applied to the center of mass of the virtual clone to simulate linear springs, is thus computed as follows:

\vspace{-3mm}
\begin{align*}
  c_c &= 2\sqrt{k m}\\
  \vec{F} &= -k\vec{d} \;+\; -c_c\vec{v}
\end{align*}

\vspace{-3mm}
\begin{itemize}[leftmargin=1.5em]
\item[$\vec{d}$] distance between the virtual clone and the real object
\item[$\vec{v}$] velocity of the virtual clone relative to the real object
\item[$k$] linear spring stiffness
\item[$m$] mass of the virtual clone
\item[$c_c$] critical damping coefficient given~$m$ and~$k$
\end{itemize}

Similarly, the torque~$\vec{\tau}$, applied to the axes of the virtual clone to simulate torsion springs, is computed as follows:

\vspace{-3mm}
\begin{align*}
  C_c &= 2\sqrt{\kappa I}\\
  \vec{\tau} &= -\kappa\vec{\theta} \;+\; -C_c\vec{\omega}
\end{align*}

\vspace{-3mm}
\begin{itemize}[leftmargin=1.5em]
\item[$\vec{\theta}$] angles (on each axis) between the clone and the real object
\item[$\vec{\omega}$] angular velocity of the virtual clone relative to the real object
\item[$\kappa$] torsional spring stiffness
\item[$I$] inertia tensor of the virtual clone
\item[$C_c$] critical damping coefficient given~$\kappa$ and~$I$
\end{itemize}

\subsection{Parameters}

From the above model, it follows that the decoupling technique is controlled by four parameters: the linear spring stiffness~$k$, the torsional spring stiffness~$\kappa$, the mass~$m$ of the object, and its inertia tensor~$I$. The inertia tensor can be computed from the mass and shape of the object. This is normally done by the physics engine.

The remaining parameters should be set according to the range of virtual masses \emph{expected to be encountered} in the virtual environment. The mass~$m$ must especially be consistent with other virtual masses. If~$m$ was much heavier than other objects, the virtual clone would have so much momentum that it would barely be affected by reaction forces during contacts, and no decoupling would occur. On the contrary, if~$m$ was much lighter, the clone would not have enough momentum to push virtual objects. Assuming virtual objects in the simulation have realistic masses consistent with their apparent size, then the mass parameter can be reasonably set to the actual mass of the real object.

The two stiffness parameters~$k$ and~$\kappa$ should be chosen to maximize the visual decoupling between light and heavy virtual objects. They should be set so that almost no decoupling occurs when pushing the lightest possible object in the simulation, and considerable decoupling occurs when pushing the heaviest possible object. Since the decoupling technique is exclusively based on this visual illusion, stiffness parameters are determinant for the effectiveness of mass perception.

An implicit parameter worth mentioning is friction. Friction is basically required in physical simulations, to prevent virtual objects from sliding indefinitely when pushed by another object. But it has also an effect on decoupling. As a force, it alters the momentum of virtual objects. Hence, according to Newton's third law, the reaction force generated when pushing on it is also altered. A closer look is necessary to ensure that this does not distort mass perception. The Coulomb model of friction states that friction depends on the mass of objects and a constant multiplication factor. The reaction force is thus increased by an amount proportional to the mass of objects. Consequently, friction should not change the fact that a virtual object is perceived as lighter or heavier than another.

\label{sec:parameters}
In our implementation, an \emph{effector object} is used to interact with virtual objects whose masses generally range from~\SI{10}{\gram} to~\SI{1000}{\gram}. We thus selected parameters to maximize the range of decoupling in this mass range. The mass parameter was set to the actual mass of the effector object:~\SI{10}{\gram}. The friction coefficient was set to~0.8 between all objects. The linear stiffness parameter~$k$ was set to \SI{50}{\newton\per\meter}, and the torsional stiffness parameter~$\kappa$ was set to \SI{150}{\newton\meter\per\radian}.

\section{Evaluation}

\subsection{Experiment 1: sorting objects by mass}

The goal of this first experiment was to confirm our base hypotheses: that pushing virtual objects with a passive real-world object (unaffected by virtual forces) does not allow to differentiate virtual masses, and that mass perception can be achieved with our proposed decoupling technique.

In this experiment, participants were using a real object called the \emph{effector} (Figure~\ref{fig:exp1-task}) to interact with virtual objects. To avoid biases in visual rendering, such as occlusion issues, we always displayed the virtual clone of the effector. In one condition, the clone was unaffected by virtual forces. In the other condition, it was decoupled as explained above. Therefore, our conditions were:
\begin{itemize}
\item[--] \textbf{C1}: the effector clone remains coupled to the real effector
\item[--] \textbf{C2}: the effector clone is decoupled from the real effector
\end{itemize}

\noindent
We were expecting~(\textbf{H1}) that condition~C1 would not allow to differentiate virtual masses. We were also expecting~(\textbf{H2}) that condition~C2 would allow to differentiate virtual masses, and specifically to correctly recognize whether a virtual object is light or heavy. Finally, we hypothesized that decoupling would not require any training to achieve mass perception~(\textbf{H3}), since it is based on a realistic metaphor. Therefore, our hypotheses were:
\begin{itemize}
\item[--] \textbf{H1}: condition~\textbf{C1} does not allow to perceive a mass difference among virtual objects
\item[--] \textbf{H2}: condition~\textbf{C2} allows to recognize which objects are lighter and heavier
\item[--] \textbf{H3}: condition~\textbf{C2} is effective with no training
\end{itemize}

\subsubsection{Participants}

Thirteen unpaid participants (3~female, 10~male) from~22 to~37 years old (mean=27.0, SD=4.9) were recruited for this experiment. One of them was left-handed. None of them had any prior knowledge of the decoupling technique.

\subsubsection{Apparatus}

Participants were seated in front of a table, on which there were the effector object, a textured AR~marker board, and a flexible stand. A~7"~tablet was attached to the stand, so that it was located between the user's eyes and the marker board. The tablet served as an augmented reality see-through display (the frame rate was 15~FPS). Virtual objects appeared on top of the marker board. The marker board and the effector object were tracked by using the Vuforia~SDK\footnote{\url{http://www.vuforia.com/}}. Participants were instructed to hold the effector behind the tablet and use it to interact with virtual objects (Figure~\ref{fig:exp1-task}).

\subsubsection{Experimental design}
\label{sec:exp1-protocol}

Prior to the experiment, participants watched a short demonstration video that showed how to hold the effector object, and how to use it to interact with virtual objects. We deemed it necessary because of the relative freedom allowed by our experimental environment. It helped participants to adopt a consistent behavior during the task. In the demonstration, the objects had exactly the same mass and there was \emph{no} decoupling. Thus, participants remained unaware of the decoupling technique up to this point.

The actual experiment started with an introductory task under the decoupling condition~(\textbf{C2}). The purpose of this first task was to evaluate the effectiveness of decoupling with fully untrained participants. Two virtual cubes, of side length~\SI{3.5}{\centi\meter}, were displayed side-by-side in augmented reality. They were visually \mbox{indistinguishable}, but had different virtual masses: \SI{15}{\gram} and~\SI{800}{\gram}. The goal was to identify the heaviest cube. Participants were asked to manipulate each cube with the effector object, and to indicate the heaviest cube by selecting it on the tablet touchscreen. The two cubes were visible at the same time, and participants were free to interact with the cubes in any order: this was intended to help participants compare the objects. There was no time limit. The task was completed as soon as the correct answer was given, but incorrect answers were also~logged.

\begin{figure}[t]
  \centering
  \includegraphics[width=.76\linewidth]{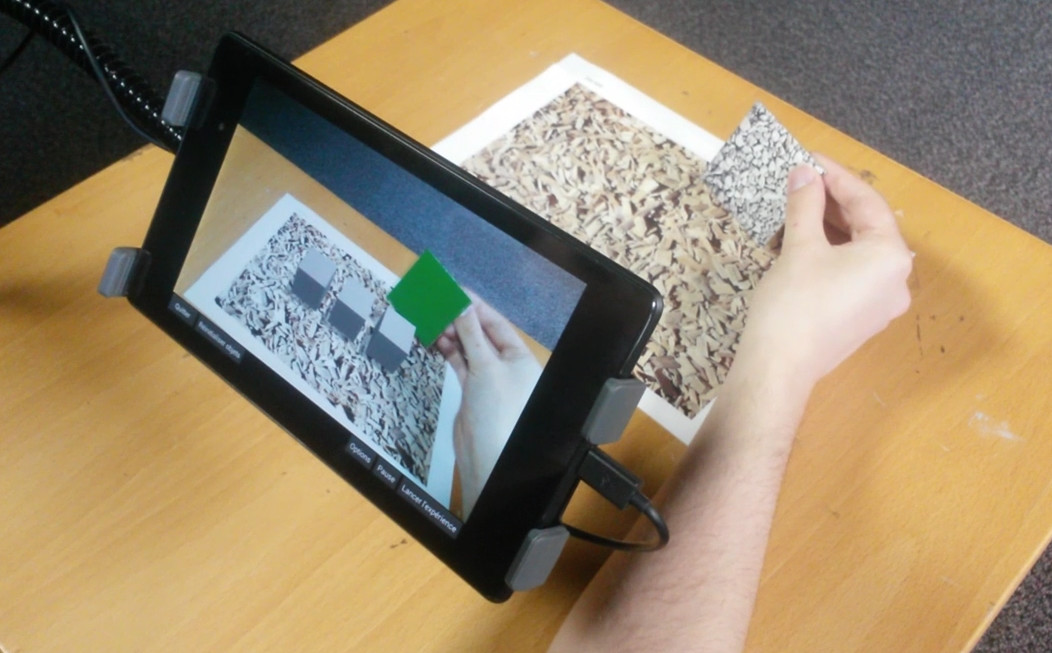}
  \vspace{-1.4mm}
  \caption{Illustration of our experimental platform. The task shown here is the sorting task of the first experiment.}
  \label{fig:exp1-task}
  \vspace{-4.5mm}
\end{figure}

After this introductory task, a series of trials started. It consisted of 10~trials: half of them were randomly assigned to each condition~\textbf{C1} or~\textbf{C2}. There were now \emph{three} virtual cubes (Figure~\ref{fig:exp1-task}), of side length~\SI{3.5}{\centi\meter}, still visually indistinguishable and with different masses. The virtual masses were~\SI{15}{\gram}, \SI{200}{\gram} and~\SI{800}{\gram}, representing typical light, intermediate and heavy objects within the supported mass range (see section~\ref{sec:parameters}). In each trial, these masses remained the same but the cubes were arranged side-by-side in a random order. The task was now to indicate which cube was the lightest, the heaviest, and the intermediate one. With 3~objects to be sorted, there was only a 1/6 chance ($\simeq$17\%) to find the right answer by pure chance. Participants were instructed to manipulate the cubes with the effector object, in any order, for as long as needed to accomplish the task. There was no time limit on each trial. At any point of the trial, participants could label a cube as the lightest, heaviest or intermediate (or correct a previous choice) by selecting it on the tablet touchscreen. Labels for each cube were mutually exclusive. After all three cubes were labeled, participants had to press a button on the touchscreen to validate their choice and complete the task. We also added a ``I~don't know'' button on the touchscreen to skip a trial. Participants were instructed to use this button in case they felt unable to accomplish the task, i.e.~when they could not perceive the masses clearly enough to tell them apart. Rather than forcing them to answer randomly, this choice ensured a more reliable result when participants themselves were certain that they did not \mbox{perceive anything}.

\subsubsection{Results and discussion}

\begin{figure}[t]
  \centering
  \includegraphics[width=.95\linewidth]{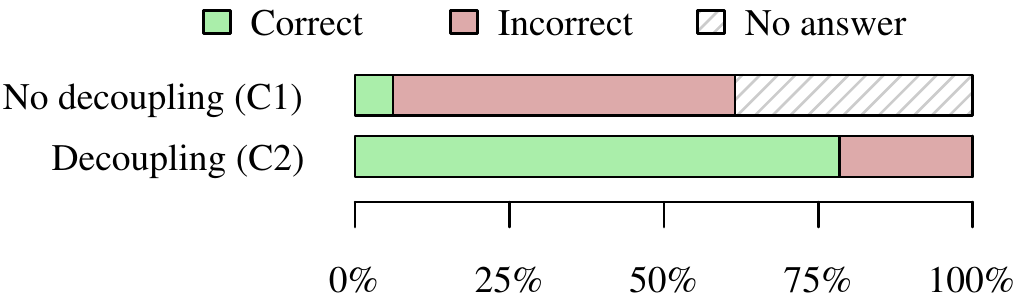}
  \vspace{-3mm}
  \caption{Mean percentage of correct, incorrect and non-answers in the sorting task. }
  \label{fig:exp1-results}
  \vspace{-4mm}
\end{figure}

Figure~\ref{fig:exp1-results} shows the results of the sorting tasks. In the~\textbf{C1} condition (no decoupling), participants were able to correctly sort the three objects in~\textbf{6.2\%} of trials on average (SD\;=\;9.6\,pp between subjects). The mean percentage of incorrect answers was~55.4\% and the mean percentage of non-answers was~38.5\%. The success rate was thus extremely low. Participants often chose not to answer, which means that they did not perceive any difference between the cubes in many cases. Even if we only consider the cases when participants actively tried to answer, the adjusted success rate~(10.1\%) is still below the 1/6~probability~(17\%) of finding a correct answer by pure chance. Therefore, these results confirm our~H1 hypothesis that condition~C1 does not allow to perceive mass differences.

In contrast, in the~\textbf{C2} condition (decoupling), participants were able to correctly sort all three objects in~\textbf{78.5\%} of trials on average (SD\;=\;20.8\,pp between subjects). The mean percentage of incorrect answers was~21.5\% and the percentage of non-answers was~0\%. This success rate was clearly above the 17\%~probability of finding a \mbox{correct} answer by pure chance. Thus, it appears that decoupling indeed allows participants to perceive a mass difference, and to differentiate between light and heavy objects. Otherwise, the success rate could not have been substantially higher than~17\%. This confirms our~H2 hypothesis. Furthermore, the high success rate shows that, at least for the decoupling parameters and virtual masses used in this experiment, participants were able to clearly and unambiguously tell the difference between objects. The fact that participants never chose to use the ``I~don't know'' button also demonstrates that they were fairly confident in their perception.

Concerning our~H3 hypothesis that decoupling is effective \mbox{without} any training, we looked at the answers to the introductory task. As explained before, even though a correct answer was required to continue the experiment, we also logged incorrect answers. However, we found that all~13~participants answered correctly on the first try. Since none of them had any exposure to the decoupling technique beforehand, we can conclude that the technique was effective with untrained subjects. This means that the decoupling metaphor can be immediately understood, which confirms our H3~hypothesis.

\subsection{Experiment 2: just noticeable difference~(JND)}

Our first experiment showed that decoupling is effective to perceive three different masses and to sort them in the correct order. However, these results were only applicable to the exact three masses tested (and for the decoupling parameters we used). Even though those three masses could be reliably differentiated, virtual masses closer to each other may be harder to distinguish.

Our second experiment was aimed at measuring the lowest mass difference that can be perceived with the decoupling technique. This measure is called ``just noticeable difference''~(JND). As a difference, a~JND is expressed relative to a reference value: in our case a \emph{reference mass}. Weber's law~\cite{ross96} states that the~JND between two real-world weights is proportional to the reference weight. Therefore, we wished to study how the~JND for decoupling changes with the reference mass. In this experiment, we measured the~JND for three reference masses within the supported mass range: \SI{15}{\gram}, \SI{200}{\gram} and~\SI{800}{\gram}. Since Weber's law was previously verified with pseudo-haptics~\cite{dominjon05}, we hypothesized that the~JND would increase when the reference mass increases~(\textbf{H1}), and that this relationship would be proportional~(\textbf{H2}):

\begin{itemize}
\item[--] \textbf{H1}: the~JND increases with the reference mass
\item[--] \textbf{H2}: the~JND is proportional to the reference mass
\end{itemize}

The same 13~participants as in the first experiment took part in the second study. This ensured they were already familiar with the decoupling technique. The apparatus was also identical (Figure~\ref{fig:exp1-task}).

\subsubsection{Experimental design}

The experimental task was to compare the mass of two objects, and to indicate whether they appeared to be different or not. In each trial, two visually identical cubes, of side length~\SI{3.5}{\centi\meter}, were displayed side-by-side in augmented reality. One of these cubes served as the reference mass, and the other as a larger comparison mass. The two cubes were arranged in a random order. Participants were instructed to manipulate the cubes with the effector object in order to compare their mass. They were asked to select the heaviest cube on the tablet touchscreen when they perceived a difference, or to press a ``No~difference'' button otherwise. There was no time limit on each trial, but participants were asked to trust their first impression and to avoid spending too much time on a single trial.

We used a staircase procedure~\cite{leek01} to estimate the~JND in relation to each reference mass. The first trial in the staircase procedure started with an initial difference between the reference mass and the comparison, and an initial step size. The values are given in Table~\ref{fig:initialdiffstep}. When a participant reported to have perceived a difference (by selecting one of the cubes), the comparison mass was decreased by the current step size.
\newcommand\tblSI[2]{\makebox[7em][c]{\makebox[4em][r]{#1}\,\makebox[3em][l]{#2}}}
\begin{table}[h]
  \vspace*{-4mm}
  \caption{Initial parameters for the staircase procedures}
  \label{fig:initialdiffstep}
  \centering
  \scalebox{0.9}{
    \begin{tabular}{ccc}
      \toprule
      \textbf{Reference mass} & \textbf{Initial difference} & \textbf{Initial step size} \\
      \midrule
      \tblSI{\textbf{15}}{\textbf{g}}  & \tblSI{175}{\gram} & \tblSI{50}{\gram} \\
      \tblSI{\textbf{200}}{\textbf{g}} & \tblSI{600}{\gram} & \tblSI{150}{\gram} \\
      \tblSI{\textbf{800}}{\textbf{g}} & \tblSI{700}{\gram} & \tblSI{200}{\gram} \\
      \bottomrule
    \end{tabular}}
  \vspace*{-1mm}
\end{table}
On the other hand, when they reported not to have perceived a difference (by pressing the button), the comparison mass was increased by the current step size. Whenever the answer was different from the previous trial, it was considered a \emph{reversal}. Every two reversals, the step size was halved. A new trial was then started with these new parameters. The full procedure ended after 10~reversals. At the end of the procedure, the~JND value was computed as the mean mass difference on all trials where a reversal occurred, except for the first two (the early range-finding trials). Since we had three reference masses, we conducted three staircase procedures in parallel by randomly interleaving trials, as a way to reduce learning effects.

\subsubsection{Results and discussion}

\begin{figure}[h]
  \vspace{-3.5mm}
  \centering
  \includegraphics[width=.95\linewidth]{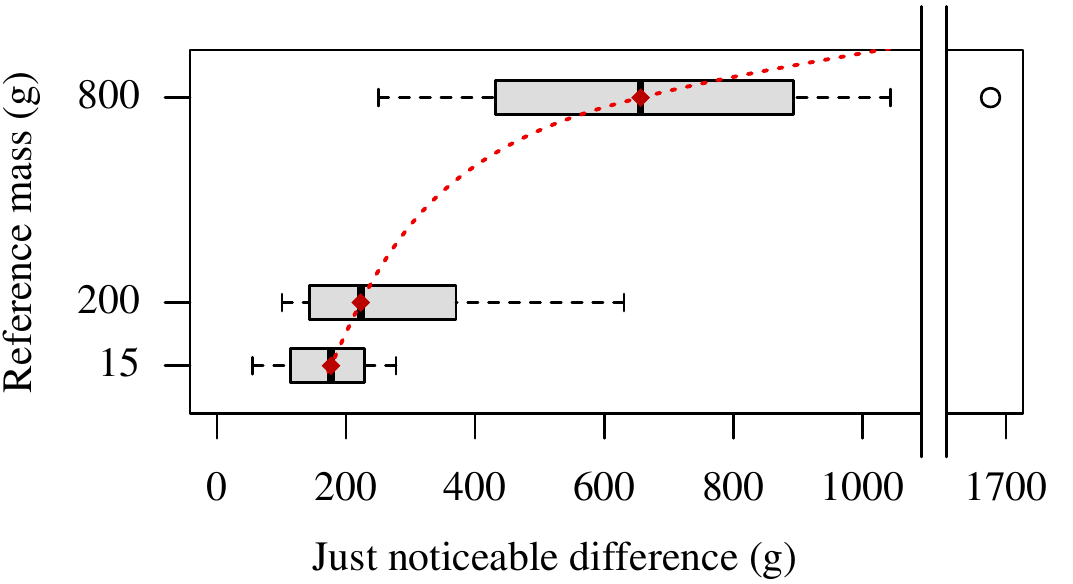}
  \vspace{-2mm}
  \caption{JND measured with respect to three reference masses. A~possible non-linear model is fitted to the three points.}
  \label{fig:jnd}
  \vspace{-.75mm}
\end{figure}

Measured~JNDs are shown in Figure~\ref{fig:jnd}. The mean values were:
\begin{itemize}
\item[--] at \makebox[2.6em][l]{\SI{15}{\gram}:}  \textbf{\SI{172.3}{\gram}} (SD=\SI{72.4}{\gram})
\item[--] at \makebox[2.6em][l]{\SI{200}{\gram}:} \textbf{\SI{271.5}{\gram}} (SD=\SI{155.3}{\gram})
\item[--] at \makebox[2.6em][l]{\SI{800}{\gram}:} \textbf{\SI{696.2}{\gram}} (SD=\SI{394.7}{\gram})
\end{itemize}

We can see that the results exhibit some dispersion. Part of it can be attributed to the inherent uncertainty of~JND measures: participants simply do not have the same perception thresholds. Another factor is the unconstrained nature of our experimental environment. For instance, participants were free to manipulate the cubes in any order. Although we instructed participants to hold the effector \mbox{object} as shown in the demonstration video (see section~\ref{sec:exp1-protocol}), they were still able to try various strategies to compare the virtual masses. \mbox{Finally,} some participants spent more time than others to get a ``first impression''. Yet, the main benefit of physics-based interaction through real-world objects is to make interaction more natural and less constrained. A highly constrained experiment would not have been realistic. By deliberately giving participants more freedom, we favored external validity of the results --~at the expense of a larger dispersion.

Nevertheless, it appears that the~JND indeed increases with the reference mass. This is consistent with previous results on mass perception, and confirms our~H1 hypothesis. Interestingly though, the relationship between the~JND and the reference mass does not seem to be proportional, as suggested by the fitting curve in \mbox{Figure~\ref{fig:jnd}}. This would invalidate our~H2 hypothesis. Obviously, three points are far from sufficient to establish a conclusive model. But this possible non-linearity could be explained by the way our decoupling technique is implemented. When interacting with a light object, such as the~\SI{15}{\gram} reference object, the force required to push it is quite small. The reaction force exerted on the virtual effector clone is thus equally small. Since the virtual clone is nearly unaffected, there is no visible decoupling. As shown in the first experiment, the lack of decoupling then makes differences between light masses unperceivable. The difference only becomes noticeable when the \mbox{reaction} force is strong enough to cause a clearly apparent decoupling, which seems to happen when the virtual mass reaches approximately 15+\SI{175}{\gram} according to the above results. This would explain why the~JND appears to have a lower bound. With heavy virtual masses, such as~\SI{800}{\gram} and above, the opposite phenomenon occurs. The force required to move the object is much higher, compared to the force generated by the spring constraint. At some point, this force \mbox{becomes} so high that the object appears to be immovable: the effector clone presses on its surface with no noticeable effect. This would explain why the~JND appears to reach a saturation point with very large virtual masses.

It must be stressed that the exact JND~values measured in this experiment remain specific to the decoupling parameters we used in this implementation. As explained before, the parameters should be tailored to maximize the decoupling differences between the lightest and the heaviest possible virtual objects. Yet, the human visual sense, the display hardware and the physical simulation all have limited precision. Therefore, adjusting parameters for a larger mass range will also worsen (increase) the~JND. On the other hand, a better (lower)~JND could be obtained by adjusting parameters to a narrower range, leading to more noticeable decoupling between two closer masses in this range.

\section{Conclusion}

In this work, we presented a new technique to visually simulate the reaction force that should occur when a real object encounters a virtual object, according to Newton's third law. This makes it possible to perceive virtual objects as having different masses. Based on pseudo-haptics, our technique does not require active force-feedback devices. In addition, unlike previous pseudo-haptics approaches in mixed reality, our technique does not require immersive visualization devices. Experimental results showed that users are immediately able to sort virtual objects by mass, with no previous training. We also measured the lowest difference~(JND) that can be perceived relative to three reference masses. Future work should focus on reducing the~JND to improve the accuracy of mass perception. Other plausible metaphors may be considered and combined. For instance, the virtual clone could be dynamically deformed when decoupling becomes too high, which would increase the visual cues when interacting with heavy objects. Other modalities, such as sound, could be exploited for a similar effect. In our current implementation, we simply superimpose the virtual clone onto the real object. Although this was shown to be effective and easy to understand, decoupling could be made more realistic with an explicit visual representation of the spring constraint (e.g.~virtual rubber bands). In its current state, however, our technique already constitutes a simple and effective way to enhance any mixed reality interface with mass perception.

\acknowledgements{This work was supported by RPE under CNRS contract 114479.}

\vspace{2mm}

\bibliographystyle{abbrv}
\bibliography{article}
\end{document}